# Quasilinear Wave "Reflection" Due to Proton Heating by an Imbalanced Turbulent Cascade

Philip A. Isenberg, Bernard J. Vasquez, Benjamin D. G. Chandran, and Peera Pongkitiwanichakul

*Institute for the Study of Earth, Oceans and Space, University of New Hampshire, Durham NH, 03824, USA*

**Abstract.** We investigate the quasilinear effects of the resonant wave-particle interaction under conditions of imbalanced turbulent heating in the collisionless coronal hole. We find that velocity-space transport of protons from the heated part of the distribution leads to strong wave growth in the minority (sunward) direction. In the present quasilinear analysis, the "reflected" waves grow to unphysical levels, indicating the necessity of including nonlinear processes. This mechanism is likely to be important in development of the fast solar wind, and may explain the puzzling minor ion observations of Landi & Cranmer (2009).

## INTRODUCTION

The plasma acceleration of the fast solar wind is a direct consequence of the high perpendicular ion temperatures in coronal holes, responding to the mirror force from the outwardly decreasing magnetic field. However, the mechanism which provides the required ion heating is still not known. The most plausible and well-developed possibility is the resonant cyclotron damping of turbulently-generated ion cyclotron waves [1]. This mechanism naturally produces the required perpendicular ion heating, including the observed preferential effects on minor ions [2-4].

The detailed feasibility of this cyclotron resonant mechanism depends entirely on the intensity and shape of the resonant spectra, which are poorly constrained both observationally and theoretically. The key question at present is whether a turbulent cascade, driven by low-frequency Alfvén waves, can transport sufficient power to cyclotron-resonant wavenumbers in the presence of the strong background magnetic field in coronal holes. MHD simulations and theory indicate that Alfvén wave power will cascade to large wavenumber primarily in the direction perpendicular to the background field, but there are still many questions and potential loopholes. Compressible interactions between Alfvén waves and fast waves have been shown to transport power to higher $k_z$ [5, 6], and quasilinear interactions with thermal protons will effectively focus oblique dispersive ion cyclotron waves into the parallel direction [7, 8]. Finally, the properties of imbalanced turbulence, where the Alfvénic fluctuations are dominated by one propagation direction and the total cross-helicity is far from zero, are also just recently under investigation.

The propagation of low-frequency MHD waves in coronal holes is primarily outward, as expected where the fundamental energy source is the Sun. This imbalanced state is observed to persist in the solar wind at least out to heliocentric radius $r \sim 9$ AU. Purely outward-propagating Alfvén waves do not evolve nonlinearly. Current theories of turbulent evolution on the open field lines of coronal holes invoke reflection of a small fraction of the Alfvén wave power to induce a nonlinear cascade. Such reflection is most effective for long wavelength oscillations, and would not result in any sunward waves on scales resonant with thermal ions. If an imbalanced turbulent cascade does extend to proton cyclotron resonant scales, those resonant fluctuations would still be propagating outward. In this case, the primary effect of cyclotron dissipation at the inner scale, would be to heat protons only in the sunward portion of the distribution, as defined by $v_z < 0$ in the plasma reference frame.

In this paper, we consider the consequences of such limited heating under the inhomogeneous conditions existing in the collisionless coronal hole. As a first step, we construct kinetic solutions of the coupled quasilinear wave-particle equations describing the interaction of protons with ion-cyclotron waves under strongly imbalanced conditions. We find that high-frequency sunward waves may be generated by the quasilinear scattering of heated protons transported across $v_z = 0$.

## QUASILINEAR EQUATIONS

The coupled quasilinear equations studied here describe the diffusion of protons in velocity space caused by resonant interactions with a broad spectrum of plasma waves and the self-consistent growth or damping of those waves due to the evolution of the proton distribution [9, 10].

The diffusion equation for a distribution $f(v_\perp, v_z, t)$ of protons with mass $m$, charge $q$, gyrofrequency $\Omega = qB/mc$ in a magnetic field $B$ in the $z$-direction is

$$\frac{\partial f}{\partial t} = \quad (1)$$

$$\frac{\pi q^2}{2m^2 v_\perp} \sum_{n=-\infty}^{\infty} \int d^3\mathbf{k}\, G\left[v_\perp \delta(\omega - k_z v_z - n\Omega)|\psi_n(\mathbf{k})|^2 G f\right]$$

where the differential operator $G \equiv (1 - k_z v_z/\omega)\,\partial/\partial v_\perp + (k_z v_\perp/\omega)\,\partial/\partial v_z$ corresponds to the gradient operator along the cyclotron-resonant surface for a wave of frequency $\omega(\mathbf{k})$ and parallel wavenumber $k_z$. The weighting function for waves of general intensity and polarization is

$$|\psi_n(\mathbf{k})|^2 = |\varepsilon_l J_{n-1} + \varepsilon_r J_{n+1} + v_z \varepsilon_z J_n / v_\perp|^2 \quad (2)$$

where $\varepsilon_l$, $\varepsilon_r$, and $\varepsilon_z$, are the fluctuating electric field components of left, right, and parallel polarization, respectively, and the argument of the Bessel functions, $J_n$, is $k_\perp v_\perp/\Omega$. The delta function in (1) specifies the resonance condition connecting protons and waves, when the Doppler-shifted wave frequency seen by a proton streaming along the magnetic field with $v_z$ is equal to an integral multiple of the proton gyrofrequency.

The corresponding equation for the intensity of the resonant waves is

$$\partial I(\mathbf{k})/\partial t = 2\gamma I(\mathbf{k}) \quad (3)$$

where the rate of change depends on the velocity-space gradients of the resonant protons

$$\gamma = \frac{\pi^2 q^2}{2m} \sum_n \frac{k_z}{\omega} \int v_\perp^2 dv_\perp dv_z \delta(\omega - k_z v_z - n\Omega) \frac{|\psi_n(\mathbf{k})|^2}{W(\mathbf{k},\omega)} G f \quad (4)$$

and $W(\mathbf{k}, \omega)$ is the total energy in the wave mode. We note that (3) is a linear equation, and requires the prior existence of a wave for any growth or damping of that mode to occur, independent of the particle gradients. We will see that this idealization limits the evolution of the system, and that nonlinear effects will have to be included to yield physically realistic results.

The self-consistent interaction with imbalanced waves needs to account for the resulting acceleration of the plasma due to the asymmetric wave heating. The bulk plasma frame, defining the wave dispersion and particle diffusion, will accelerate outward along the magnetic field. This acceleration, $dU/dt$, creates a backward inertial force on the protons, changing (1) to

$$\frac{\partial f}{\partial t} - \frac{dU}{dt}\frac{\partial f}{\partial v_z} = \left(\frac{\partial f}{\partial t}\right)_{w-p} \quad (5)$$

where the right-hand side of (5) denotes the wave-particle diffusion of (1).

## MODEL ASSUMPTIONS AND COMPUTATIONAL PROCEDURES

This initial investigation considers a spatially homogeneous system evolving in time. Some of the effects studied are caused by the inhomogeneous expansion of the plasma in a coronal hole, but here we will approximate these effects within a uniform space to avoid adding more variables to the calculation.

The wave properties are described by the cold plasma dispersion relations for an electron-proton plasma, in the limit of zero electron mass. In this limit $\varepsilon_z \to 0$. Further, high-order resonances require large parallel proton speeds, which are absent in the low-$\beta$ plasma of the coronal hole. Therefore, we neglect the terms in (1) and (4) when $n \neq 1$, leaving only the ion-cyclotron wave mode.

Oblique waves are included here, but we restrict the propagation angles to $\theta \leq 45°$. Highly oblique waves on this dispersion branch become kinetic Alfvén waves, whose primary interactions are with electrons [11], and these are not treated realistically here. We will also see that the dominant effects here concern waves at small angles to the field.

The velocity space of the protons and the $\mathbf{k}$-space of the waves are each defined on 2-D grids. The ($v_\perp$, $v_z$) grid is centered in the rest frame of the bulk protons and taken to be uniform in each variable. The boundary conditions are the usual reflective condition at $v_\perp = 0$, and absorbing conditions at the other boundaries, placed far enough from the origin to yield no significant loss of particles. The $(k, \theta)$ grid is uniform in $\theta$. The $k_z$ grid values are related to the $v_z$ grid values through the resonance condition, and the uniform $\theta$ grid then determines $k$. The linear growth rate (4) depends only on local values of the proton gradients, so boundary conditions are not needed. In the calculations here, the $v_z$ grid has 1200 values between $\pm 0.2\, V_A$, the $v_\perp$ grid has 400 values up to $0.4\, V_A$, and the $\theta$ grid has one-degree spacing.

The initial proton distribution is taken to be Maxwellian, with a thermal speed $v_{th}^2 = 10^{-3}\, V_A^2$. The outward-propagating wave spectrum is a 3D Kolmogorov power law

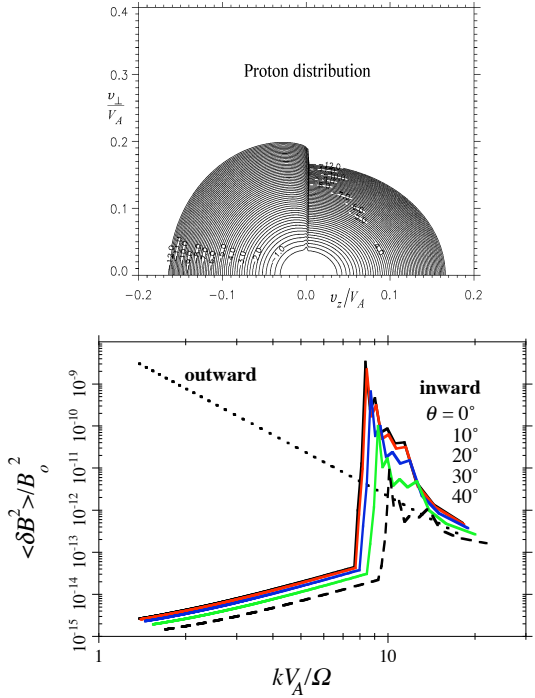

**FIGURE 1.** Soln. of the homogeneous case at $t = 5\times10^5/\pi\Omega$. (Top) Contours of the proton distribution in the plasma frame. (Bottom) Wave intensities in the outward direction (dotted line) and five angles in the inward direction.

$$I_{out}(k, \theta) = 10^{-8} (kV_A/\Omega)^{-11/3} B_o^2 \qquad (6)$$

for $\theta \leq 45°$, constant in time. This relative intensity is consistent with the level of wave power at the proton inertial scale used in our models [3, 4] at $r = 2\, R_S$. This level is a small fraction of the outward resonant wave power extrapolated from observations. These models assumed nearly balanced wave spectra, and imbalanced conditions in a coronal hole may allow much larger outward intensities. We do not address waves at angles $\theta > 45°$ in this paper.

Equation (3) requires the presence of "seed" waves for any interaction to proceed. We take the initial sunward waves to scale with the smallest intensity of outward waves in the calculation, at $k_{max}(\theta)$, where $k_{max}$ is the wavenumber resonant with the smallest $v_z$ on the grid. We then set the seed wave electric fields constant in $k$: $<\delta E^2>_{in}(k, \theta, t=0) = <\delta E^2>_{out}(k_{max}, \theta)$. The seed wave magnetic field is then a function of $k$, as determined by the dispersion relation. For the velocity grid used here, $k_{max} \sim 16\, \Omega/V_A$, which corresponds to $I_{in} < 3\times10^{-13}\, B_o^2$.

We alternatively advance (3) and (5) in time, taking the acceleration from the imbalanced heating to be a constant during each time step. After each time step, the bulk plasma speed in the accelerated frame is calculated from the new proton distribution, and the value of the acceleration constant is iterated to maintain $<v_z> \sim 0$ in this frame.

## RESULTS

As expected, the steady spectrum of outward-propagating resonant waves heats the sunward protons for all cases. The resulting bulk acceleration creates a net proton transport in the sunward direction with respect to the plasma rest frame, such that the proton density for $v_z < 0$ exceeds that for $v_z > 0$ by a fraction of a percent. These protons slide sunward primarily from the very-low-energy core near $v_\perp = 0$, leaving a small trough in the anti-sunward distribution there.

Sunward-propagating waves will be generated if sufficient heated protons are transported back across $v_z = 0$ at high $v_\perp$. However, the anti-sunward protons, essentially still in Maxwellian form, strongly damp any low-level sunward waves. In this initial-value problem, the sunward seed waves immediately disappear and thus the proton diffusion for $v_z > 0$ vanishes as well. The sunward protons continue to be heated by the steady outward waves, but any wave-particle interactions for $v_z > 0$ are confined to the first $v_z$ grid point and are of questionable validity. The anti-sunward protons are basically unaffected.

A more realistic scenario is obtained by declaring that the non-ideal microscopic processes responsible for the initial seed waves will persist in maintaining them in time. We approximate this case by imposing a "floor" on the level of sunward waves, such that $<\delta E^2>_{in}(k, \theta) \geq <\delta E^2>_{out}(k_{max}, \theta)$ for all times, rather than just initially. In this case, the seed waves allow for some proton diffusion across $v_z = 0$, representing the effects of noisy small-scale phenomena, such as compressive fluctuations or collisions, on the low $v_z$ particles at the center of the distribution. The results, shown in Fig. 1 for $t = 5\times10^5/\pi\Omega$, are striking. On the top is a contour plot of the proton phase-space density, clearly showing the heating of sunward protons by the steady outward waves and a small amount of spillover into the $v_z > 0$ region. On the bottom, the magnetic spectra of sunward waves at five angles to the background magnetic field are shown, along with the imposed outward wave spectrum for reference. The smooth curves for $k < 7\, \Omega/V_A$ indicate the magnetic intensities corresponding to the floor minimum values taken here. We see that the slight excursion of heated protons across $v_z = 0$ causes a huge enhancement of $\sim 10^5$ in the small-scale sunward wave spectra, concentrated in the parallel direction. However, the interaction here remains a boundary phenomenon, confined to small $v_z$ and large $k$, without causing any significant heating of the anti-sunward protons.

Some improvement is made by noting that protons in coronal holes are subject to gravity and the mirror force, which increase transport across $v_z = 0$. These forces arise from the inhomogeneity of the medium, but rather than add a spatial variable to the calculation, we use uniform forces in the kinetic equation (5) that approximate these effects at 2 $R_S$ in a coronal hole. A fully inhomogeneous calculation will be presented in future work. The mirror force acceleration is set proportional to $v_\perp^2$, with the proportionality constant estimated from the coronal hole model of [3]. The gravitational acceleration of the Sun is standard. Figure 2 shows the results of this case at $t = 7\times10^5/\pi\Omega$. We see that the heated protons manage to penetrate into the anti-sunward region to a somewhat greater extent than in the previous case, but these enhancements are still narrowly confined. Similarly, the wave spectra have peaks an order of magnitude larger than the previous case, but they only extend to slightly lower $k$ and still cut off abruptly before reaching more moderate wavenumbers.

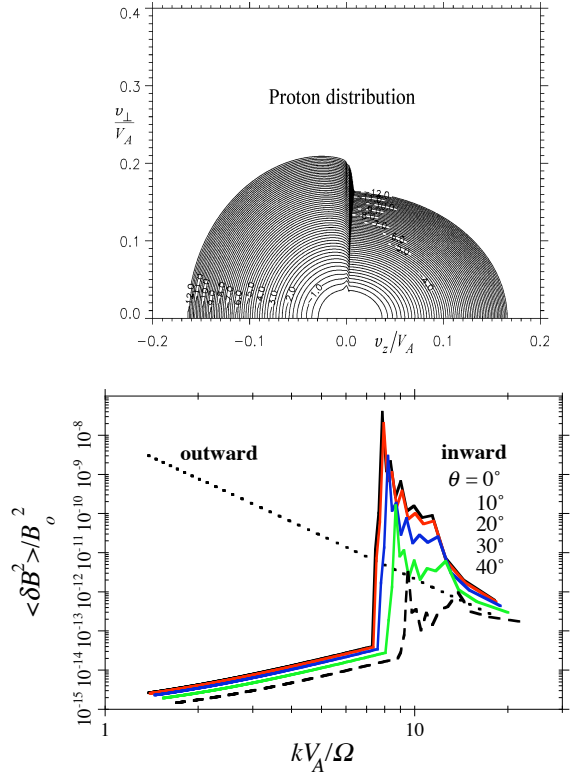

**FIGURE 2.** Solution of the "inhomogeneous" case at $t = 7\times10^5/\pi\Omega$. (Top) Contours of the proton distribution in the plasma frame. (Bottom) Wave intensities in the outward direction and five angles in the inward direction.

Unfortunately, our numerical calculation breaks down at times later than shown here, so we cannot yet make definitive statements on the eventual heating of anti-sunward protons by this process. The numerical difficulties result from the extreme gradients in the solutions, which would not exist if nonlinear wave transport processes were included. Realistically, some form of energy transport in **k**-space will act to limit wave power enhancements long before they reach the $10^6$ levels. The next step in this model will be to include such transport, in the form of resonance broadening or a diffusion in wavenumber [12-15].

## CONCLUSIONS

We have shown that cyclotron resonant heating by strongly imbalanced waves will lead to unstable proton distributions in the inhomogeneous conditions typical of the collisionless coronal hole. The instability causes intense linear growth of the minority (sunward) wave intensity at cyclotron resonant wavenumbers, which we term "quasilinear reflection". This process is distinct from the true reflection of outward waves by gradients in the Alfvén speed, which occurs in the opposite limit of very small $k$, and is thought to drive the MHD turbulent cascade in the corona. Addition of nonlinear wave interactions should result in an inverse transport of resonant wave power, which could explain the peculiar charge/mass dependence of minor ions reported by [16].

## ACKNOWLEDGMENTS

This work was supported in part by the NSF Space Weather Program, grant ATM0719738, the NSF SHINE Program, grant ATM0851005, NASA grants NNX07AP65G and NNX08AH52G, and DoE grant DEFG0207ER46372 to the Center for Integrated Computation and Analysis of Reconnection and Turbulence.